\documentclass[aps,prl,twocolumn,showpacs,preprintnumbers,amsmath,amssymb,superscriptaddress]{revtex4}
\usepackage{graphicx}
\usepackage{dcolumn}
\usepackage{bm}

\begin{document}

\title{Enhancement of the  $\nu = 5/2$ Fractional Quantum Hall State
in a Small In-Plane Magnetic Field}

\author{Guangtong Liu}
\affiliation{Department of Electrical Engineering, Princeton
University, Princeton, New Jersey 08544}

\author{Chi Zhang}
\affiliation{Department of Electrical Engineering, Princeton
University, Princeton, New Jersey 08544}
\affiliation{Department
of Physics and Astronomy, Rice University, Houston, Texas
77251-1892}

\author{D. C. Tsui}
\affiliation{Department of Electrical Engineering, Princeton
University, Princeton, New Jersey 08544}

\author{Ivan Knez}
\affiliation{Department of Physics and Astronomy, Rice University,
Houston, Texas 77251-1892}

\author{Aaron Levine}
\affiliation{Department of Physics and Astronomy, Rice University,
Houston, Texas 77251-1892}

\author{R. R. Du}
\affiliation{Department of Physics and
Astronomy, Rice University, Houston, Texas 77251-1892}

\author{L. N. Pfeiffer}
\author{K. W. West}
\affiliation{Department of Electrical Engineering, Princeton
University, Princeton, New Jersey 08544}

\date{\today}

\begin{abstract}
Using a 50-nm width, ultra-clean GaAs/AlGaAs quantum well, we have
studied the Landau level filling factor $\nu = 5/2$ fractional
quantum Hall effect in a perpendicular magnetic field $B \sim$ 1.7 T
and determined its dependence on tilted magnetic fields. Contrary to
all previous results, the $5/2$ resistance minimum and the Hall
plateau are found to strengthen continuously under an increasing
tilt angle $0 < \theta  < 25^\circ$ (corresponding to an in-plane
magnetic field 0 $<$ $B_\parallel$ $< 0.8$ T). In the same range of
$\theta$ the activation gaps of both the $7/3$ and the $8/3$ states
are found to increase with tilt. The $5/2$ state transforms into a
compressible Fermi liquid upon tilt angle $\theta
> 60^\circ$, and the composite fermion series [2+$p/(2p\pm1)$], $p
=$ 1, 2 can be identified. Based on our results, we discuss the
relevance of a Skyrmion spin texture at $\nu = 5/2$ associated with
small Zeeman energy in wide quantum wells, as proposed by
W$\acute{\text o}$js $et$ $al$., Phys. Rev. Lett. 104, 086801
(2010).
\end{abstract}

\pacs{73.43.Qt, 73.43.Nq, 73.21.Fg}
\keywords{fractional quantum Hall state, in-plane magnetic field,
} \maketitle

The fractional quantum Hall effect (FQHE) observed at Landau level
(LL) filling factor $\nu =5/2$ \cite{1,2,3} has received much
attention since theory suggests that its quasiparticles may obey
non-Abelian statistics \cite{4,5}. Here $\nu= n_eh/eB$, where $n_e$
is the sheet density of 2D electrons, and $B$ is the magnetic field.
To date, most of the evidence supporting the non-Abelian nature of
$5/2$ is from exact diagonalization or numerical calculations
\cite{6,7} based on modeling electron-electron interaction
potentials (including the Coulomb potential), although more recently
experimental evidence has accumulated supporting quarter-charged
quasiparticles at $5/2$ \cite{8,9,10}, or non-Abelian statistics
\cite{10}. If the predicted non-Abelian properties can be firmly
established both theoretically and experimentally, the braiding of
these non-Abelian particles can form the basis for topologically
protected quantum computation \cite{11}.

\begin{figure}
\includegraphics[width=0.8\linewidth]{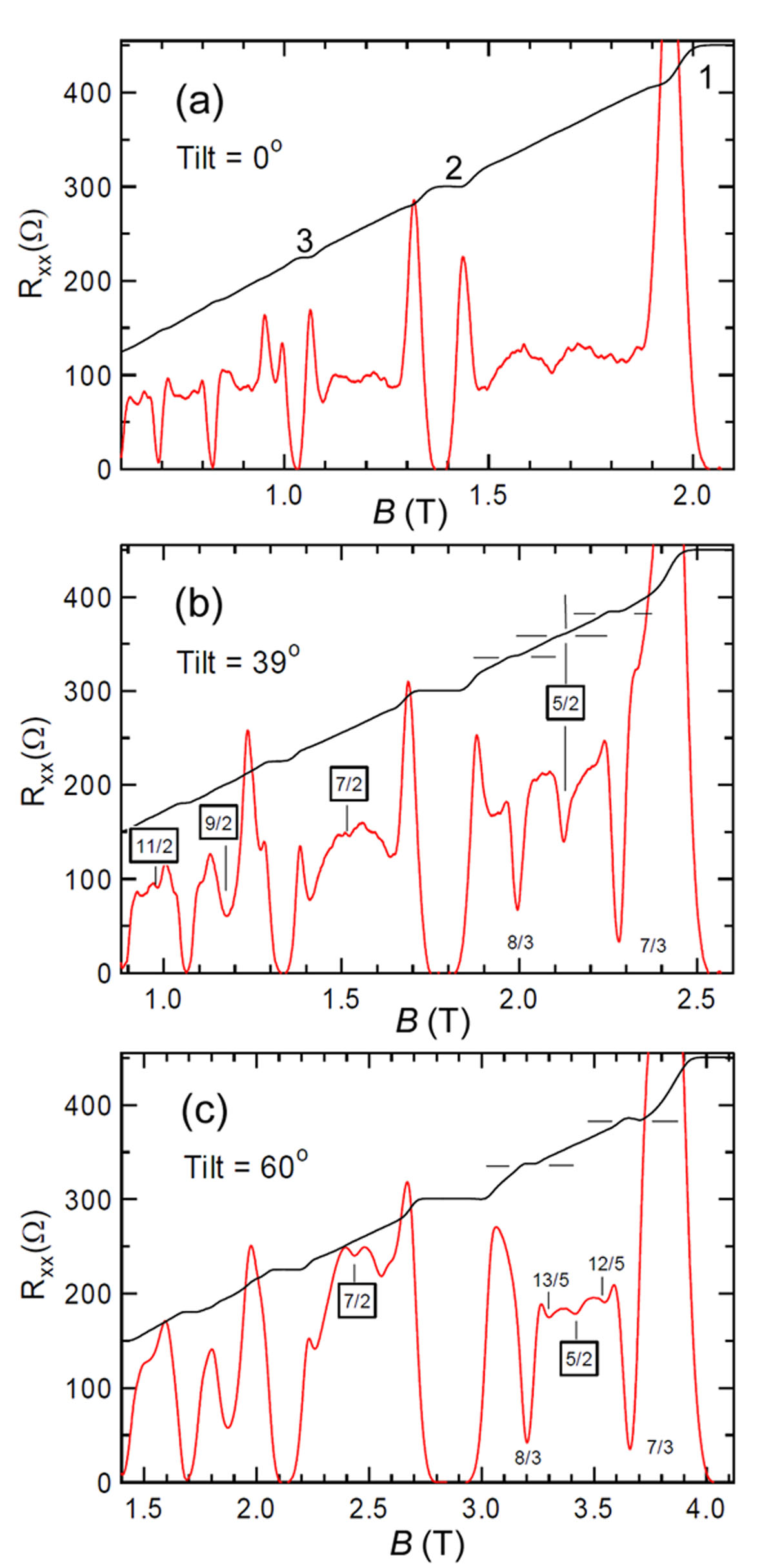}
\caption{\label{fig:fig1} {(color online) Sample A: the
magnetoresistance (R$_{xx}$) and Hall resistance (R$_{xy}$) traces
are shown for representative tilt angles: (a) $\theta  = 0^\circ$,
R$_{xx} (5/2)$ shows a cusp but no discernable plateau in R$_{xy}$.
No resistance features for the $7/3$ and $8/3$ FQHE were observed;
(b) $\theta  = 39^\circ$, R$_{xx}$ ($5/2$) develops into a minimum
associated with a visible Hall plateau in R$_{xy}$. Strong FQHE
states at both the $7/3$ and $8/3$ filling factors are observed; (c)
$\theta  = 60^\circ$, the states around $5/2$ develop into a
composite fermion state series at $\nu =2+p/(2p \pm 1)$, $p =$ 1, 2.
All traces were measured at a temperature $\sim$ 25 mK.}}
\end{figure}

The proposed Moore-Read (MR) Pfaffian wavefunction for $5/2$
requires the spin being at least partially polarized. Recent
numerical results indicate that the $5/2$ state in a GaAs quantum
well system is spin-polarized even in the limit of vanishing Zeeman
energy \cite{12}. Experimental investigations on spin polarization
at $5/2$ from electrical transport \cite{13,14,15,16} and optical
measurements \cite{17,18,19} have been reported, but these results
provide only indirect determination of the spin states. Recent
resistively detected nuclear magnetic resonance measurements point
to a fully spin-polarized ground state (GS) at $5/2$ at a magnetic
field around 5 T \cite{20,21}.

It is anticipated that increasing the Zeeman energy would help to
stabilize the spin-polarized GS in the presence of fluctuations.
Therefore, a tilted magnetic field is supposed to enhance the FQHE
at $5/2$ \cite{12}. However, to date all the experimental results
have contradicted this simple prediction: the $5/2$ FQHE is found to
be weakened in a tilted field \cite{13,14,15,16}. Competition with a
striped many-electron phase could be a plausible cause for the
complex response of the $5/2$ state in a tilted field
\cite{22,23,24}. On the other hand, W$\acute{\text o}$js $et$ $al$.
point to an interesting possibility based on numerical calculations
in wide GaAs$/$AlGaAs quantum wells (QWs), the MR state could be
depolarized by forming a Skyrmion - spin texture or spin topological
defects \cite{25,26}. Moreover, Skyrmions are further promoted by
disorder \cite{25}. It appears then for realistic samples, the GS at
$5/2$ could experience a spin texture to fully polarized phase
transitions as the Zeeman energy is increased. In high $n_e$
samples, Zhang $et$ $al$. \cite{15} reported a quantized $5/2$
plateau at 10 T, which is taken as the evidence for a polarized GS.
For the opposite limit of small Zeeman energy, we report in this
Letter a $5/2$ FQHE observed at 1.7 T in a low $n_e$ sample, where
the $5/2$ plateau and energy gap are found to enhance by a small
in-plane magnetic field. Our results render strong experimental
support for the notion that the true ground state at $5/2$ must be
spin-polarized; in realistic materials, the Zeeman energy stabilizes
the ground state against fluctuations.

Our samples were modulation-doped symmetrical square
Al$_{0.24}$Ga$_{0.76}$As$/$GaAs$/$Al$_{0.24}$Ga$_{0.76}$As QW with a
well width of $w = 50$ nm and a spacer width of $d = 220$ nm. The
specimens, A and B, were approximately 3 mm $\times$ 3 mm square
each with eight In$/$Sn ohmic contacts diffused around the
perimeter. After a brief red light-emitting diode (LED)
illumination, the specimen attained a mobility  $\mu \sim$ 1.5
$\times$ 10$^7$ cm$^2$$/$Vs at 300 mK and an electron density $n_e =
1.00 \times 10^{11}/$cm$^2$ (sample A) and 1.02 $\times
10^{11}/$cm$^2$ (sample B). The magnetotransport experiments were
performed in a top-loading dilution refrigerator with a base
temperature of 20 mK and an 18-T superconducting magnet in the
National High Magnetic Field Laboratory (NHMFL). Tilt-field
measurements were facilitated by an $in$ $situ$ sample rotator, with
the total magnetic field $B_{total}$ applied with an angle $\theta$
with respect to the sample normal. Standard lock-in technique (7 Hz
and 2 nA excitation current) was employed for resistance
measurements.

The central findings from our tilted magnetic field experiments in
the 50 nm QW were: 1) the $5/2$ FQHE was enhanced under a small
in-plane magnetic field $B_\parallel$; and 2) the $5/2$ FQHE
transitioned to a composite fermion state in larger $B_\parallel$.
The representative data taken from sample A are shown in Fig. 1.
Here the R$_{xx}$ and R$_{xy}$ traces were measured at $T \sim$ 25
mK. At $\theta  = 0^\circ$, we observed a well-defined cusp in
R$_{xx}$ at $5/2$ (magnetic field $B \sim$ 1.66 T), but no Hall
plateau was discernable at this temperature. As can be seen in Fig.
3(a), for sample B at $T\sim$ 20 mK, a weak R$_{xx}($5/2$)$ minimum
and a developing R$_{xy}($5/2$)$ plateau could already be observed
at $\theta  = 0^\circ$. In any case, the $5/2$ FQHE state in this QW
at zero tilt angle appears to be rather fragile and the strength of
the R$_{xx}($5/2$)$ minimum is sensitively dependent on the sample
cooling and LED illumination conditions \cite{27}.

\begin{figure}
\includegraphics[width=0.9\linewidth]{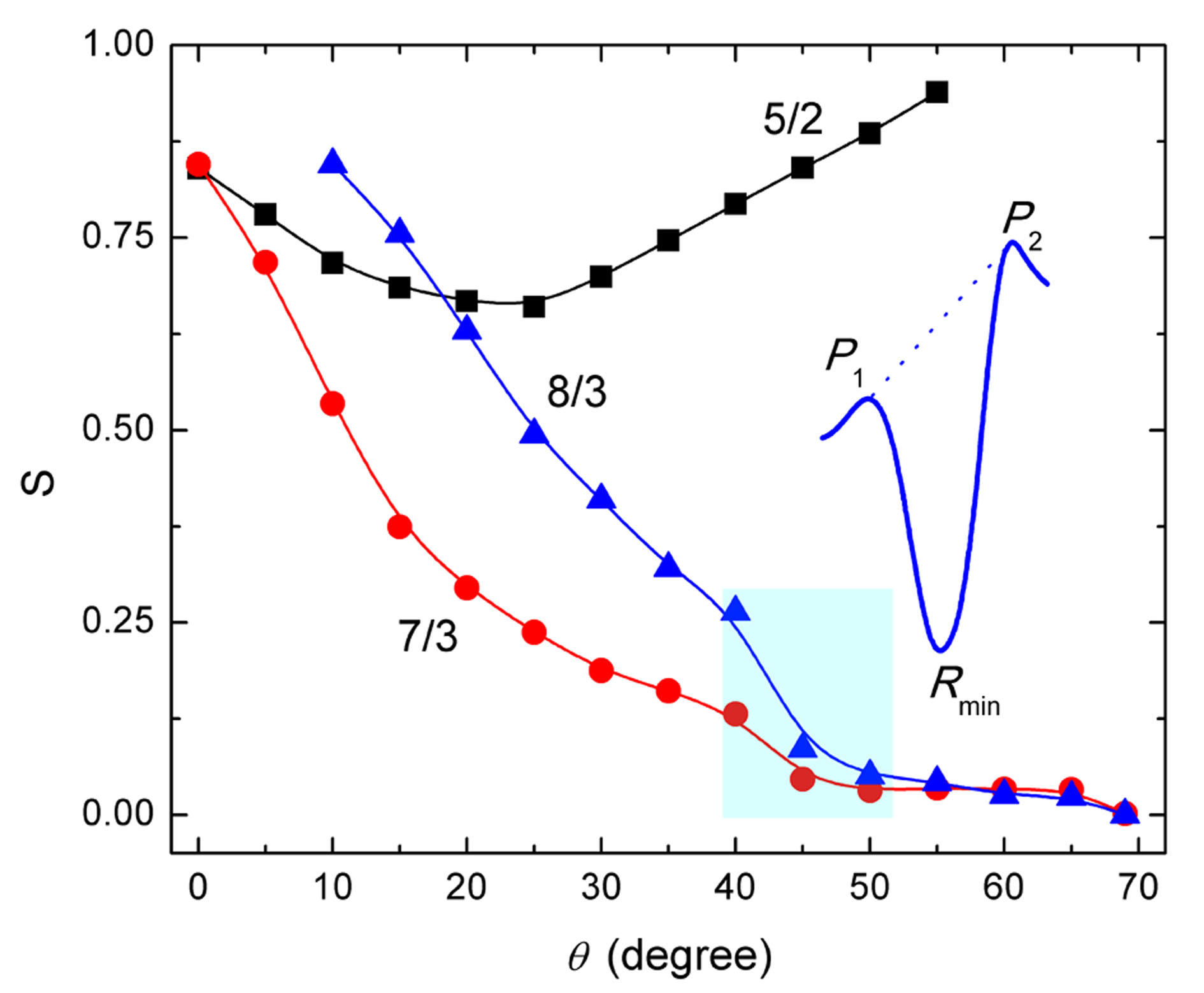}
\caption{\label{fig:fig2} {(color online) Sample B: the evolution of
parameter S as a function of tilt angle $\theta$ is shown for $5/2$,
$7/3$, and $8/3$. The solid curves are a guide to the eye. The
schematic in inset shows the peak-valley structure which defines S
(see text). The square area indicates a steep enhancement of the
$7/3$ and the $8/3$ FQHE in the tilt angle range $\sim$ 40$^\circ -
50^\circ$, as discussed in the text.}}
\end{figure}

Both sample A and sample B show quantitatively the same response to
a tilted magnetic field. Surprisingly, even a small tilt angle
$\theta \sim 10^\circ$ yielding an in-plane magnetic field
$B_\parallel \sim$ 0.3 T dramatically enhanced the $5/2$ FQHE. The
R$_{xx}($5/2$)$ minimum deepened, along with the development of an
R$_{xy}($5/2$)$ plateau in a continuous fashion for tilt angles $0 <
\theta < 25^\circ$. Above these angles, the trend reverses. Figure
1(b) shows an example for the FQHE at $\theta$=$39^\circ$, where a
Hall plateau at R$_{xy}$ = ($2/5$) ($h/e^2$) is visible, along with
a deep minimum in R$_{xx}$. We also observe that at this tilt angle,
the peak around the R$_{xx}($5/2$)$ becomes very steep, indicating
that the tilt-induced insulating phases (IP) becomes relevant in
$\theta \geq 40^\circ$. It is important to note that both the $7/3$
and the $8/3$ FQHE states, which are not discernable at $\theta =
0^\circ$ at this temperature ($\sim$ 25 mK), have become remarkably
strong in tilted fields, as evidenced by quantized Hall plateaus and
steep resistance minima for both $\theta \sim 39^\circ$ and $\theta
\sim 60^\circ$.

\begin{figure}
\includegraphics[width=0.9\linewidth]{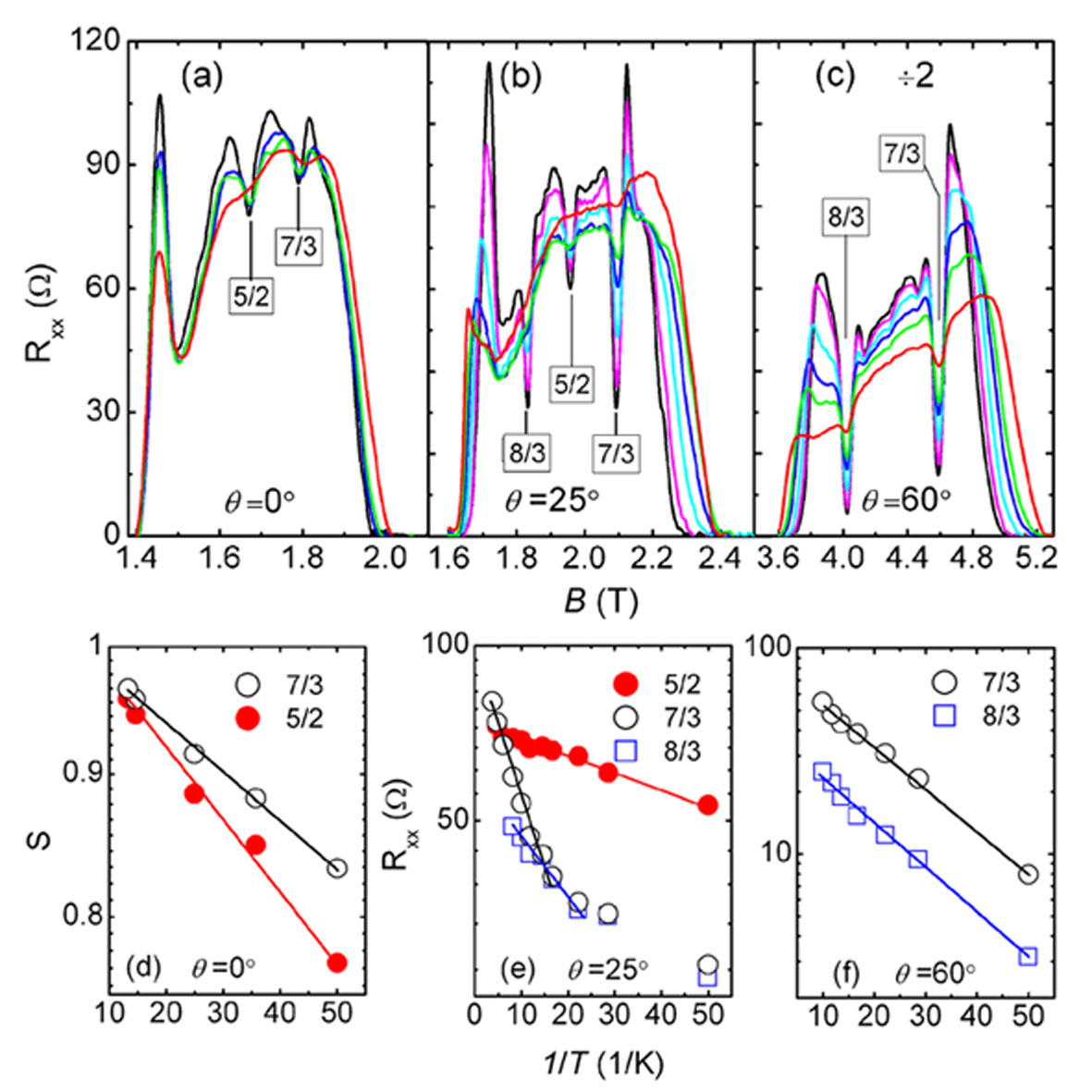}
\caption{\label{fig:fig3} {(color) Sample B: (a)-(c)
Temperature-dependent magnetoresistances R$_{xx}$. The data were
taken at temperatures of 20mK, 28mK, 40mK, and 75mK in (a); 20mK,
45mK, 85mK, 123mK, 166mK, and 266mK in (b); 20mK, 45mK, 74mK, 101mK,
123mK, and 208mK in (c). (d)-(f) Corresponding S (defined in the
text) versus $1/T$, or R$_{xx}$ versus $1/T$ at filling factor
$5/2$, $7/3$, and $8/3$ are shown respectively for tilt angle (d)
$\theta = 0^\circ$: $\Delta^\text S_{5/2} \approx$ 10 mK,
$\Delta^\text S_{7/3} \approx$ 7 mK; (e) $\theta = 25^\circ$:
$\Delta_{5/2} \approx$ 13 mK, $\Delta_{7/3} \approx 106$ mK,
$\Delta_{8/3} \approx$ 46 mK; (f) $\theta = 60^\circ$: $\Delta_{7/3}
\approx$ 88 mK, $\Delta_{8/3} \approx 103$ mK.}}
\end{figure}

For a true FQHE state, it is expected that its resistance minimum
shows temperature-activated behavior, namely, R$_{xx}$ $\propto$
exp($-\Delta/{2k_BT}$), where $\Delta$ is the gap energy and $k_B$
is the Boltzmann constant. The robustness of the FQHE is measured by
$\Delta$. However, a developing FQHE state often shows only limited
range of activation, and therefore $\Delta$ is not experimentally
accessible. In order to quantitatively assess the strength of the
FQHE at $5/2$, $7/3$, and $8/3$ in a tilted field, we plotted the
ratio of resistance minimum and the average peak value (see the
inset of Fig. 2), S$\equiv$2R$_{min}/(P_1+P_2)$, against tilt angle.
A similar procedure to estimate the $5/2$ strength was used in Refs.
\cite{13,14,15}. The data were taken from sample B at the base
temperature $T =$ 20 mK. Such plots exhibit systematic patterns for
the relevant FQHE states in a wide range of $0^\circ < \theta <
70^\circ$.

Two types of distinctively different patterns are observed: 1) For
$7/3$ and $8/3$, the S values decrease monotonically with the tilt
angles, indicating a continuous strengthening of the FQHE at these
filling factors. The result for $7/3$ is consistent with that
previously reported by Dean $et$ $al$. \cite{13} for a QW having a
40 nm width and an electron density of $n_e \sim 1.6\times
10^{11}/$cm$^2$. The initial $7/3$ state ($\theta \sim 0^\circ$) in
\cite{13} is proposed to be a Skyrmion, based on the fact that the
ratio of the Zeeman energy and the electron interaction energy $\eta
\sim$0.01 is within the range of the critical ratio for Skyrmion
formation near $\nu =$1. 2) For $5/2$, S first decreases, reaching a
broad minimum at $\theta \sim 25^\circ$, and then increases. This
pattern is in sharp contrast with that reported for a 40 nm QW
\cite{13}, where S($5/2$) monotonically increases with tilt angle.

We note that in previous tilt-field experiments, in-plane magnetic
fields consistently enhance the IP, which are revealed as resistance
spikes around $5/2$ \cite{22,23,24}. Here this effect appears to be
weak in the tilt angle range $\theta \sim 30^\circ$. It is plausible
that the competition between the FHQE liquid at $5/2$ and the IP
around it, which cannot be completely resolved in realistic samples,
has led to a complex response of resistance in this filling factor
range, and that only in the present samples, the quantum liquid
characteristics at $5/2$ prevails. On a side note, the precipitous
drop of S($7/3$) and S($8/3$) between $\theta \sim 40^\circ$ and
$\sim 50^\circ$, as marked by the square area in Fig. 2, is not
understood; it could be due to the enhancement of IP in this range,
or, a precursor to the transition into a composite fermion (CF)
series [2$+p/(2p \pm 1)$], with $p =$ 1, 2.

Xia $et$ $al$. reported \cite{16} a tilt-induced anisotropic to
isotropic phase transition at $\nu =5/2$, observed in a 40 nm QW.
For increasing tilt angles, the authors found that the anisotropic
compressible phase can be replaced by isotropic compressible phases
reminiscent of the CF fluid at $\nu =1/2$. In our 50 nm QWs, the
electrical subband separation is calculated to be $\sim$ 5.8 meV
\cite{28,29,30}, which indicates that subband mixing in the presence
of an in-plane magnetic field is significant. We thus believe that
the observed CF series at high tilt angles is a result of subband
mixing, confirming the explanation in \cite{16}.

In Fig. 3 we plot the temperature-dependent magnetoresistance for
the tilt angles $\theta \sim 0^\circ$, 25$^\circ$, and $60^\circ$,
respectively, measured from sample B. We note that for $\theta$
$\sim 25^\circ$, R$_{xx}($5/2$)$ clearly shows dR$_{xx}/$d$T$ $>$ 0,
as expected for a developing quantum liquid. This is in contrast to
the case of $\theta \sim 0^\circ$, where dR$_{xx}/$d$T$ $\sim$ 0.
Such plots demonstrate that, indeed, the $5/2$ FQHE is being
enhanced by a small in-plane magnetic field component $B_\parallel
\sim$ 0.7 T in this QW. In order to quantitatively access the energy
scale of the relevant states, we determined the energy gap
$\Delta^\text S$ from S $\propto$ exp($-\Delta^\text S/2k_BT$), see
Fig. 3(d). For the FQHE states at $5/2$, $7/3$, and $8/3$ in tilted
fields, we determined the activation gap $\Delta$ by fitting the
data to an Arrhenius relation, R$_{xx} \propto$
exp(-$\Delta/2k_BT)$. The values of $\Delta$ and $\Delta^\text S$
are presented in the caption of Fig. 3.

In summary, we have observed an enhancement of the low magnetic
field FQHE at filling factor $5/2$ in a small in-plane magnetic
field. This finding is in stark contrast to all previously reported
results. In the following paragraphs we discuss briefly the possible
relevance of our observation to the phase transition from a
partially-polarized to a fully spin-polarized ground state at $5/2$
driven by Zeeman energy. This regime has not been explored
experimentally until this work.

For the $5/2$ state in a perpendicular magnetic field $B$, the ratio
between the Zeeman energy $E_Z=g \mu_B B$ and the electron
interaction energy $E_C=e^2/\varepsilon l_B$, $\eta \equiv E_Z/E_C$
scales as $\sqrt{B}$, where $g$ is the effective $g$ factor in GaAs,
$\mu_B$ is the Bohr magneton, $\varepsilon$ is the dielectric
constant, and $l_B=\sqrt{\hbar/{eB}}$ is the magnetic length.
W$\acute{\text o}$js $et$ $al$. point out that a wide QW favors the
formation of Skyrmions, because the Coulomb repulsion for a
spin-singlet pair of electrons softens considerably (relative to a
spin-triplet) upon increasing the well width. At a sufficiently low
Zeeman energy, it is energetically favorable for charge $e/4$
quasiholes to bind into Skyrmions of charge $e/2$ \cite{25}. In
small $B \sim$ 1.7 T, $\eta \sim$ 0.008, which is comparable to the
critical value of $\sim$ 0.01 for forming quasiparticle spin
textures, as calculated in \cite{26}. In realistic samples,
potential fluctuations are always present and these can further
promote Skyrmions by bringing two repelling quasiparticles together.
Based on these theoretical considerations, it is plausible that the
observed weak plateau at 5$/$2 and its enhancement in a small tilt
angle may be relevant to a Skyrmion spin texture at $\nu = 5/2$ in
the limit of small Zeeman energy in wide quantum wells.

In order to further examine the origin of these tilt-field effects,
we analyzed quantitatively the behavior of the 5$/$2 state and the
nearby 7$/$3 and 8$/$3 states in small tilt angle range $\theta
<$25$^\circ$. General properties of the Skyrmion states in the FQHE
regime have been studied both theoretically [see, e.g.,
\cite{31,32,33}] and experimentally [see, e.g., \cite{34,35}].
Characteristically, such states prevail in the limit of a small
Zeeman energy (originating from a small $g-$factor or a small
applied magnetic field). In transport measurements the activation
energy at the FQHE filling factor as a function of Zeeman energy can
be assessed to reveal the existence of Skyrmions \cite{34,35}. In
the Skyrmion state the gap increases faster than a single spin-flip
upon increasing the total magnetic field. In the case of $''$valley
Skyrmions$''$, it is found that the interaction-induced quantum Hall
gap increases faster than the single particle energy \cite{36}.

Estimated from our data in the small tilt angle range, the $7/3$ gap
increases faster than single spin-flip energy, while for $8/3$ and
$5/2$ the gap openings are better described by a single spin-flip.
For the lack of experimental $g$-factor values at these fractions,
here we take the electron bare $g$-factor in GaAs, which is a
reasonable choice based on the fact that composite fermions carry
the same $g$-factor \cite{37,38}. This observation appears to
contradict Skyrmion properties, suggesting that a different ground
state may be relevant at zero tilt. On the other hand, the Skyrmion
model proposed for 5$/$2 involves fractionally-charged
quasi-particles, and how Zeeman energy alters the many-body energy
gap remains an open question \cite{25,26}, which should be addressed
in future work. We note that in \cite{38}, the energy gap of 8$/$3
in very high-mobility GaAs$/$AlGaAs QWs of different electron
density (hence different magnetic field) have been carefully
measured, and the results can be clearly explained by a Landau fan
diagram of composite fermions carrying a spin on the second Landau
level. Altogether, spin properties of the FQHE states in the second
Landau level appear remarkably rich and remain only partially
understood.


We acknowledge very helpful discussions with Wei Pan, Li Lu, and
Changli Yang. R.R.D. thanks Srinivas Raghu for discussions
concerning 331 state versus MR state at different magnetic fields.
The work at Princeton was partially funded by the Gordon and Betty
Moore Foundation as well as the National Science Foundation MRSEC
Program through the Princeton Center for Complex Materials
(DMR-0819860). The work at Rice was funded by DOE
(DE-FG02-06ER46274). A portion of this work was performed at the
NHMFL, which is supported by NSF Cooperative Agreement No.
DMR-0084173, by the State of Florida, and by the DOE.


\begin{references}
\bibitem{27} In order to prepare samples showing the well-developed FQHE states, a slow cool-down procedure was followed; typically it took $\sim$2 hours
             to cool sample from 300 K to 2 K. A red LED with a bias current of 10 mA was used to illuminate the sample at 2K for 30 minutes. We note
             that such procedure was standard in experiments on very high mobiliy GaAs$/$AlGaAs quantum wells (see, e. g., Ref\cite{3}).
\bibitem{28} The electrical subband separation is calculated based on a self-consistent solution of the Schrodinger and Poisson equations.
             The band bending effect \cite{29,30} due to electron distribution has been taken into account.
\end{references}
\end{document}